      [1999/12/01 v1.4c Il Nuovo Cimento]
\documentclass{cimento}

\usepackage{graphicx}  
\title{Recent progress in QCD calculations for $e^{+}e^{-}$ annihilation and hadron collisions}
\author{M.~Dasgupta\from{ins:x}\ETC,
}
\instlist{\inst{ins:x} School of Physics and Astronomy, The University of Manchester, Oxford Road, Manchester M13 9PL, United Kingdom.}
\begin{document}

\maketitle

\begin{abstract}
We provide a brief summary of recent developments in QCD calculations in and 
beyond fixed-order perturbation theory for observables in $e^{+}e^{-}$ annihilation as well as hadron collisions.
\end{abstract}

\section{Introduction}
The interplay between physics at hadron colliders and that at 
$e^{+}e^{-}$ machines has traditionally been of great significance in 
furthering our understanding of elementary particles and their interactions. 
One can for example point to the specific case of the discovery of the $Z$ boson at a hadron collider \cite{Z} which was then followed by 
high precision phenomenology at LEP which helped to establish firmly the standard model of particle physics, the current theory of elementary particles beyond which any discoveries are still to be made. 

This tradition is set to continue with the strong expectation that the 
LHC will lead to the discovery of the Higgs boson or help to clarify the Higgs sector as well as enabling the discovery of physics beyond the standard model. The extremely high energy hadronic collisions at the LHC, which make it a 
powerful discovery machine, however come with a price which takes the form of a more complicated initial state (protons rather than elementary particles) and complications concerning non-perturbative effects such as beam remnant interactions (the underlying event) and pile-up which threaten to limit the 
theoretical precision that one may be able to obtain. The most precise 
determination of the parameters of the new physics such as masses and couplings would probably require a cleaner environment such as a high energy 
$e^{+}e^{-}$ future linear collider. 

Nevertheless as the Tevatron experience has to an extent confirmed, 
calculations in perturbative QCD will have a strong role to play in the physics program of the LHC, particularly with regards to estimating accurately backgrounds to new physics. To this end significant effort has been devoted in the past years to develop QCD calculations specifically for important LHC processes in the discovery context. Moreover given the vast scale hierarchy inherent in high energy hadron collider physics (with scales ranging from the TeV range centre-of--mass energy $\sqrt{s}$, through typical jet transverse momenta $p_T$, the masses of electroweak scale particles down to the few GeV scales associated to non-perturbative physics) it is clear that techniques involving summation of 
large logarithms in scale ratios would also be important in maximising the theoretical accuracy one may be able to achieve. The introduction of new and faster infrared and collinear (IRC) safe jet algorithms and a systematic 
understanding of perturbative and non-perturbative properties of jets and jet substructure is also a rapidly developing and vital part of the current and future LHC physics program. 

At the same time, as should be clear from the preceding discussion, 
furthering the precision of QCD calculations for $e^{+}e^{-}$ annihilation
remains of continued importance for future phenomenology as well as remaining a simpler learning and testing ground for QCD practitioners. In this context 
the development of next-to--next to leading order (NNLO) predictions and taking resummed computations from the state of the art next-to--leading logarithmic (NLL) level through to NNLL accuracy as well as possibly improving the current theoretical understanding of non-perturbative effects such as hadronisation 
corrections will all play an important role.

In what follows below we present a brief summary of what we perceive to be some of the main developments and recent progress in QCD calculations for both 
hadron colliders and $e^{+}e^{-}$ machines. It is impossible due to page 
limitations to adequately cover all the relevant progress that has been made in the past few years and thus the selection of topics/references below is far 
from complete. We shall aim to discuss briefly the progress in fixed-order 
perturbative computations as well as all-order resummations both in the hadron collider and the $e^{+}e^{-}$ context, mention the status of $\alpha_s$ measurements and discuss progress in the definition and understanding of jets and their properties in and beyond QCD perturbation theory.

\section{QCD at fixed order}
Observables that have the property of infrared and collinear (IRC) safety can be calculated as an expansion in the strong coupling $\alpha_s$ using perturbative techniques based on the evaluation of Feynman graphs. By an IRC safe observable one essentially means the following: Let ${\mathcal{O}}_n \equiv {\mathcal{O}}(p_1,p_2,\cdots p_n)$ denote the value of the observable $\mathcal{O}$ due 
to a configuration involving $n$ partons with momenta $p_1,p_2 \cdots p_n$.
Now consider adding an extra parton with momentum $p_{n+1}$. In the soft limit that the energy $E_{n+1} \to 0$ (with $E_1, \cdots ,E_n$ held finite) or the limit that $\vec{p}_{n+1} \to \vec{p}_i$ where $i=1,\cdots,n$ i.e the limit in which the three-momentum $\vec{p}_{n+1}$ is parallel to any of the three-momenta $\vec{p}_i$ (collinear limit) IRC safety implies that independent of $n$, $\mathcal{O}_{n+1} \to \mathcal{O}_n$. IRC safety ensures that real-virtual cancellation of divergences occurs and hence that finite results are obtained 
in perturbation theory. 

For a simple observable of the above kind, $V$, involving a single hard scale $Q^2$, we can then write the perturbation expansion as 
\begin{equation}
V = \sum_{n=0}^{\infty} C_n \left(\frac{Q^2}{\mu^2} \right) \alpha_s^n(\mu^2),
\end{equation}
where $C_n$ are perturbatively calculable coefficients, $Q$ is the hard scale of the process and $\mu$ an arbitrary renormalisation scale, which however should be chosen to be of order $Q$ to avoid large logarithms in $Q^2/\mu^2$. The dependence on $\mu$ would in fact cancel if one were able to compute the observable to all orders exactly but in practice one is able to evaluate only a few terms in the above sum. The residual $\mu$ dependence in a calculation truncated at $n^{\mathrm{th}}$ order in $\alpha_s$ is of the order of uncalculated ${\mathcal{O}}\left(\alpha_s^{n+1}\right)$ terms. Thus scale dependence is usually 
taken as a measure of the influence of uncalculated higher orders and hence the theoretical accuracy of a given prediction. \footnote{One should be aware that the scale dependence may in cases not be a reliable estimate of the true size of higher orders. For example if new hard scattering channels open up in higher perturbative orders varying scales in a lower order contribution cannot be expected to estimate the size of such new contributions.} 
 
Generally speaking leading order (LO) calculations are too crude to be considered reliable estimates for most collider observables. NLO calculations on the 
other hand may be expected to be correct, broadly speaking, to within order 10 percent while NNLO calculations represent high precision and as a rule of thumb ought to be accurate to within a few percent or so.
\footnote{There are exceptions to these broad statements which for instance only apply to observables not afflicted by multiple disparate hard scales. For 
explicit counter examples see for instance Ref.~\cite{giant}.}
\begin{figure}
\includegraphics[width=0.6\textwidth]{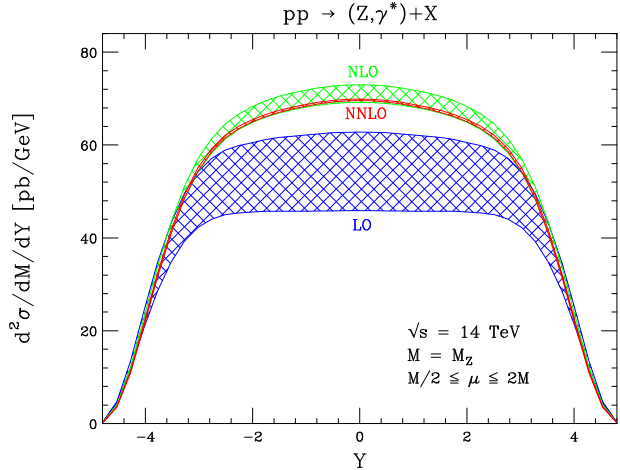}   
\caption{An illustration of the scale uncertainty reduction with the order of 
the perturbative estimate for the case of the rapidity ($Y$) distribution for 
inclusive $Z$ production at the LHC. Figure taken from Ref.~\cite{Babis}.}
 \label{fig:babz}
\end{figure}
An illustration of this is provided in Fig.~ 1 where one notes the progressive reduction in scale uncertainty with the increasing order of the perturbative evaluation for the case of the inclusive $Z$ rapidity distribution for the LHC. 

For reliable estimates of backgrounds to LHC processes with new physics it 
would thus appear that at least NLO accuracy is a must. For up to the production of three jets at hadron colliders NLO calculations encoded in the program 
NLOjet++ have been available for some time \cite{ref:Nagy}.
However many of the relevant discovery processes involve high multiplicity final states with similar backgrounds involving e.g multiple hard final state jets for which it is much less straightforward to obtain NLO estimates. At present the current state of the art for NLO computations at hadron colliders is for $2 \to 4$ processes such as a $t \bar{t} b \bar{b}$ final state relevant for 
Higgs production and decay in association with a $t \bar{t}$ pair \cite{Shitmeir1,Shitmeir2}. Similarly NLO calculations to $W+3j$ \cite{Shitmeir3,Shitmeir4} and $Z^0+3j$ \cite{Shitmeir5} have been recently computed. A significant development in the computation of NLO corrections has been the advent of unitarity based calculational methods alongside traditional Feynman-diagram techniques. A pedagogical review and further references can be found in Ref.~\cite{GZ}. The 
automation of NLO computations is also an important step towards the calculation of several different collider processes. The automation of both real radiation terms \cite{SHERPA,MadGraph,Phegas} and virtual corrections \cite{Golem,BlackHat,CutTools,Rocket,Samurai} has been achieved in the past few years, for NLO corrections. 

As far as NNLO calculations are concerned only a few processes are known to 
such accuracy. For instance for hadron collisons fully exclusive NNLO corrections to vector boson production have been computed \cite{Catani,Petriello} 
while for the case of $e^{+}e^{-}$ annihilation similar calculations have been performed using the method of antenna subtraction for the case of $e^{+}e^{-} \to 3j$ which has enabled a more accurate determination of $\alpha_s$ from data on LEP event shape variables \cite{Glover,thefreak,Bethke}.

Having briefly summarised the state of the art for QCD calculations at fixed-order we shall turn our attention to those observables where the involvement of more than one perturbative scale forces us to go beyond fixed-order 
perturbation theory using resummation methods.

\section{QCD beyond fixed-order}

As mentioned above there exist several observables of phenomenological interest where multiple scales (typically the process hard scale and other scales introduced due to observable definition) play an important role. For such 
observables, the classic examples of which remain event or jet shape variable 
distributions \cite{Dassal}, one has to consider the role of large logarithms 
in scale ratios and examine the possibility to resum these to all orders at a 
given logarithmic accuracy.
 
To be more explicit consider the distribution in some shape variable $\tau$ in say $e^{+}e^{-}$ annihilation:
\begin{equation}
\frac{1}{\sigma} \frac{d\sigma}{d\tau} \sim \sum_n \frac{1}{\tau} \alpha_s^n \ln^{2n-1} \frac{1}{\tau}+\cdots
\end{equation}

The above behavior reflects the double-logarithmic enhancement of the shape 
cross-section due to soft and collinear gluon emissions while the ellipsis denote less singular terms some of which also need to be accounted for for phenomenological purposes. This result is clearly divergent and 
unphysical at small $\tau$ which reflects the inadequacy of fixed-order 
predictions in that region and hence the need for resummation. 

On resummation, for those variables that have the property of exponentiation 
\cite{CTTW} one can write a result of the form
\begin{equation}
\frac{1}{\sigma} \frac{d\sigma}{d\tau} \sim \frac{d}{d\tau} e^{-C_F \alpha_s \ln^2 \frac{1}{\tau}}+\cdots
\end{equation}
which generalises with account of running coupling and less singular terms into the form ($L \equiv \ln 1/\tau$):
\begin{equation}
\frac{1}{\sigma}\frac{d\sigma}{d\tau}\sim \frac{d}{d\tau} \exp \left [Lg_1(\alpha_s L) +g_2(\alpha_s L) +\alpha_s g_3 (\alpha_s L) +\cdots \right].
\end{equation}

In the above result the leading and next-to--leading logarithmic (NLL) terms are represented by the functions $g_1$ and $g_2$. The current state-of--the art 
for most observables at any collider process is NLL accuracy in the resummed exponent. The NNLL function $g_3$ is known only for some select variables 
amongst which are the thrust and heavy jet-mass distribution in $e^{+}e^{-}$ annihilation (in fact computed in the framework of soft-collinear effective theory to $\mathrm{N^3 LL}$ accuracy \cite{BecNeu}) and for hadron collisions the Drell-Yan and Higgs transverse momentum ($Q_T$) distribution (see for instance 
Ref.~\cite{CatNLL} and references therein). Most recently for the Drell-Yan case results have also been obtained including NNLL accuracy for the new $a_T$ and $\phi^*$ variables measured by the D0 collaboration \cite{d0} which broadly speaking are in good agreement with the data even without inclusion of non-perturbative effects \cite{at,DasMarBan}.

For $e^{+}e^{-}$ annihilation the role of resummation in 
ensuring precision phenomenology has been clear for a long time \cite{CTTW}. 
\begin{figure}
\label{fig:scet}
\includegraphics[width=0.4 \textwidth]{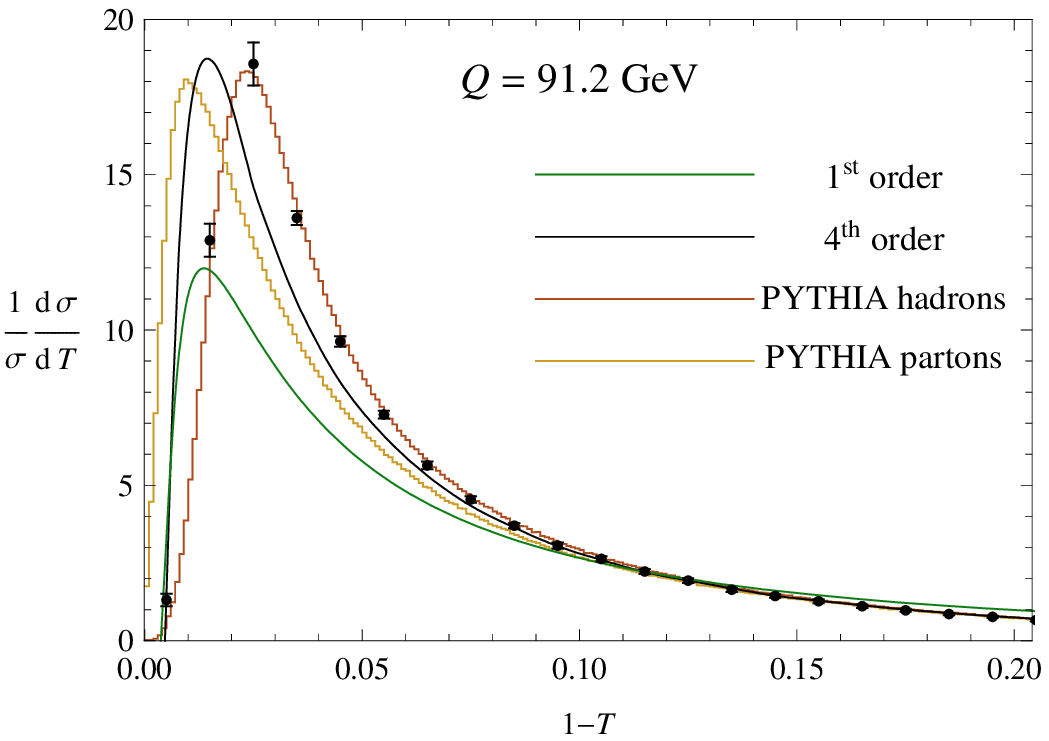}
\includegraphics[width=0.4 \textwidth]{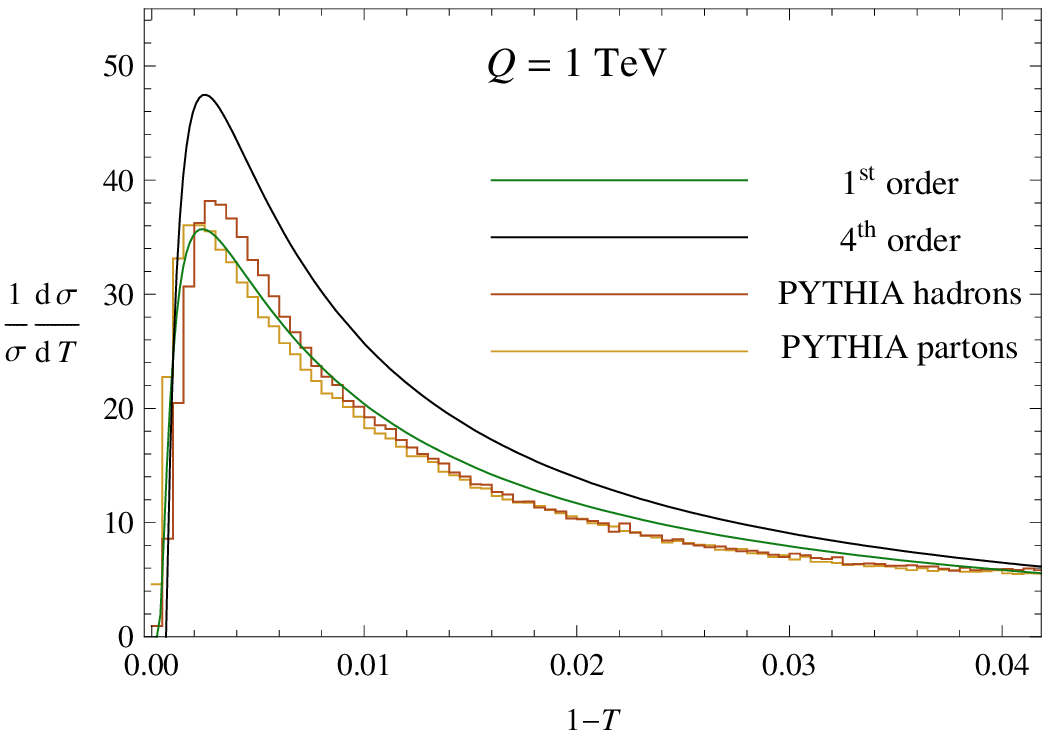}
\caption{Figure illustrating the comparison between various levels of resummation and results from PYTHIA for the thrust distribution in $e^{+}e^{-}$ 
annihilation for $Q=91.2$ GeV (left) and $Q=1$ TeV (right). Data from LEP are also shown in the former case. Figure taken from Ref.~\cite{BecNeu}}
\end{figure}
Consider as a recent example the comparison between various levels of resummation, event generators and $e^{+}e^{-}$ event shape data depicted in Fig.~2. 
At the $Z$ peak it appears that there is excellent agreement between PYTHIA (at hadron level) and data. Moreover the PYTHIA (parton level) result appears rather closer to the $\mathrm{N^3LL}$ ($4^{th}$ order) curve than to the LL result which is where one may expect it to be. That this is an effect which arises 
due to uncontrolled sub-leading terms and the tuning procedure inherent in 
PYTHIA is revealed by going to $Q=1$ TeV, where for example subleading effects would be inconsequential, PYTHIA is much closer to the LL rather than the $\mathrm{N^3LL}$ result. It has hence been observed in Ref.~\cite{BecNeu} that using LL MC generators may potentially lead to a significant underestimate of certain QCD backgrounds at a future ILC (at about the few tens of percent level).
\begin{figure}
\label{fig:cae}
\includegraphics[width =0.8 \textwidth]{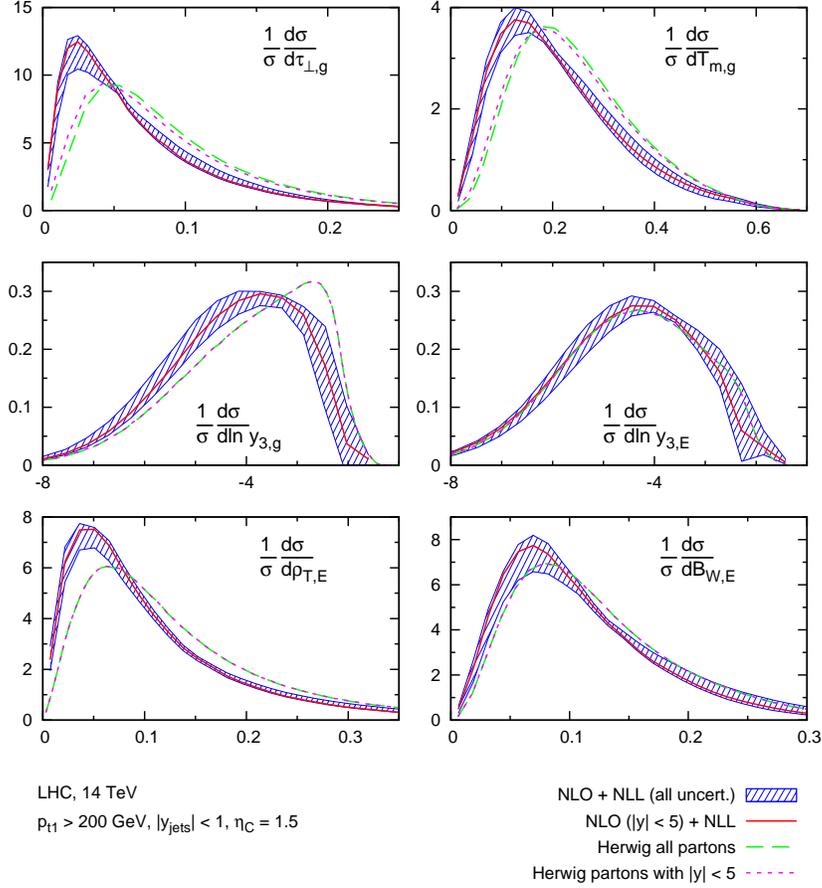}
\caption{Resummed predictions for global hadron collider event shapes compared to results from HERWIG. Figure taken from Ref.~\cite{BSZ}}
\end{figure}

While accurate resummed predictions have been an important requirement in say the determination of $\alpha_s$ from LEP event shapes, they are also in principle of great value for jet production in hadron collisions in terms of improving perturbative accuracy. However the more complex hadronic environment at a hadron collider makes all-order resummation a rather delicate affair. For instance care has to be taken in constructing observables such as event shapes to avoid contamination from beam remnants by constructing suitably central event shapes which then have the property of being non-global \cite{DassalNG1,DassalNG2}. Since the non-global single logarithms cannot be computed beyond the large $N_c$ limit, in order to ensure full NLL accuracy for observables such as  event 
shapes in hadronic dijet production, the observables have to be further 
modified in such a way so as to ensure globalness, such as those variables 
studied in Ref.~\cite{BSZ}. A yet more troublesome issue is the contamination as a result of effects such as pile-up which can potentially override the eventual accuracy which can be achieved via theoretical methods such as resummation. It is thus desirable to seek variables that are less prone to such effects in order to test resummed calculations hadron collider observables. Predictions for several hadron collider event shape variables as reported in Ref.~\cite{BSZ} are shown in Fig.~3. In some cases some discrepancy with corresponding results from leading-log and leading colour event generators such as HERWIG can also be noted. For more detailed comments on the role of tuning and the shower parameters in such comparisons we refer the reader to the comments in Ref.~\cite{BSZ}. Detailed phenomenological studies for hadron collider event shape variables are currently in progress \cite{BSZ,Aaltonen}. 

Aside from a limited number of global event shapes and observables such as suitably defined dijet azimuthal correlations \cite{BanDasDel} as well as Drell-Yan $Q_T$ spectra, one may try to study via resummation other observables involving for instance jet-definition and the application of a jet algorithm. As an 
example of this one can point to the case of jet masses and shapes for high $p_T$ jets at the LHC which are relevant in identifying the origin of a jet as being initiated by a QCD process (quark or gluon jet) or say by the decay of a boosted heavy particle. The QCD jet mass distribution for example would receive logarithmic enhancements $\sim \alpha_s \ln^2 \frac{R P_t}{M_j}$ where $P_t$ is the transverse momentum and $M_j$ the jet mass, with $R$ being the jet radius. Since at the LHC we will encounter jets with $P_t$ in the TeV range, the role of such logarithmic terms can be expected to be substantial even up to jet masses near the electroweak scale. The resummation of such logarithms while being 
immensely desirable from the standpoint of perturbative accuracy however has 
complex issues mainly to do with the role of non-global logarithms and 
jet algorithms and was recently discussed in Ref.~\cite{DasMarBanKam}. While a very high formal level of precision in such cases is essentially ruled out it 
should still be possible to develop resummation formulae that capture the numerically dominant terms in the result to sufficient accuracy for phenomenological purposes.

We conclude this section by presenting in Fig.~4, the current status of $\alpha_s$ determinations taken from Ref.~\cite{bethke}. The 2009 value for the world average for $\alpha_s(M_z)$ was reported as $0.1184 \pm 0.0007$. The individual contributions from various QCD observables used for $\alpha_s$ extraction are also shown in Fig.~4.
\begin{figure}
\label{fig:asbet}
\includegraphics[width=0.5\textwidth]{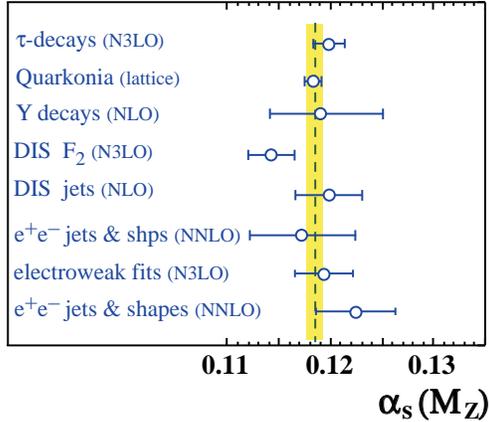}
\caption{Figure displaying the $\alpha_s$ values extracted from various QCD 
studies alongwith the world average value (dashed vertical line) and error (yellow band). Taken from the comprehensive 2009 review \cite{bethke}.}
\end{figure}

\section{Progress in jet definiton and understanding jet properties}
Although the precise definition of QCD jets may appear a detail not necessarily directly relevant to progress of high order QCD calculations discussed in the major part of this review, it is in fact the case that such calculational developments need to be supported by suitable IRC safe jet definitions. In other words higher order perturbative estimates for jet cross-sections and differential distributions only make sense when an infrared and collinear safe jet algorithm is used in jet definition. Although in many cases of interest such as inclusive jet $p_T$ spectra the IRC unsafety of a given jet algorithms may only appear at a relatively high order, for several LHC processes involving large multiplicity of final state jets (say as backgrounds to a new physics process) the IRC unsafety may appear already at leading order invalidating any level of perturbative accuracy \cite{SalSoy}. 
Likewise it is not meaningful to compute all-order resummed predictions for quantities that will diverge at any fixed order due to the algorithm in use. This requirement coupled with experimental and practical considerations (speed of the algorithm for high multiplicity hadronic final states) make the definition of jets a non-trivial task. Fortunately there now exist several different practically feasible options for IRC safe jet definitions defined either using sequential recombination \cite{k_t,C_A,anti} or based on the idea of cone jets \cite{SalSoy}. The recent fast progress in the field of jet physics are expertly reviewed in Ref.~\cite{Salam} to which we point the interested reader for further details.

As a by product of the rapid developments in jet physics there has also recently been tremendous interest in using a somewhat more sophisticated understanding of jets and their properties, gained via relatively simple analytical 
calculations, as a chisel for improving the prominence of new physics signals at the LHC. For example the idea of the optimal value of jet radius $R$ to be employed in various searches for new physics at the LHC based on analytical 
estimates of both perturbative radiation and non-perturbative hadronisation 
corrections was suggested in Ref.~\cite{Dassalmag}. 

Moreover ideas about jet substructure \cite{Sey,ButtSal} have contributed to an explosion in the production of tools which can be used to distinguish QCD 
jets from those produced by the decays of massive particles in the highly 
boosted regime where the decay products may be captured within a single jet. 
For a detailed exposition of substructure techniques we refer the reader to Ref.~\cite{boost} and references therein. 

To conclude we finish with a reminder that much of the progress in developing 
QCD precision tools and the consequent improvement in understanding QCD effects whether in the context of hadron colliders or $e^{+}e^{-}$ machines should 
ultimately yield benefits beyond the particular context within which it was initiated. For instance the need for developing theoretical methods to further the precision that can be obtained via perturbative techniques at the LHC should 
in many cases ultimately have spin-offs that would pay dividend in the attainment of even higher precision at future linear colliders. There is thus much 
reason to be optimistic in light of the fact that the pace of developments of 
QCD tools continues to be rapid (and possibly even accelerated) stimulated in 
large part by the advent of LHC data.

\end{document}